\documentclass[11pt, a4paper]{article}
\usepackage{jheparxiv}
\usepackage[latin1]{inputenc}
\usepackage{amsmath}
\usepackage{amsfonts}
\usepackage{amssymb}
\usepackage{latexsym}
\usepackage{mathrsfs}
\usepackage{graphicx}
\usepackage{color}
\usepackage{slashed}
\usepackage{twistor}

\subheader{\hfill \texttt{DAMTP-2014-53}}

\title{A Worldsheet Theory for Supergravity}

\author{Tim Adamo, Eduardo Casali and David Skinner}

\affiliation{Department of Applied Mathematics \& Theoretical Physics \\
        University of Cambridge \\
        Wilberforce Road \\
        Cambridge CB3 0WA, United Kingdom}

\emailAdd{[t.adamo, e.casali, d.b.skinner]@damtp.cam.ac.uk}

\abstract{We present a worldsheet theory that describes maps into a curved target space equipped with a $B$-field and dilaton. The conditions for the theory to be consistent at the quantum level can be computed exactly, and are that the target space fields obey the nonlinear $d=10$ supergravity equations of motion, with no higher curvature terms. The path integral is constrained to obey a generalization of the scattering equations to curved space. Remarkably, the supergravity field equations emerge as quantum corrections to these curved space scattering equations.}

\begin{document}
\notoc

\maketitle\vfill

\section{Introduction}

It is a highly non-trivial (if well-known) fact that General Relativity emerges as the low energy limit of closed string theory.  This equivalence was first observed via the tree-level S-matrices of the two theories: the $\alpha^\prime\rightarrow 0$ limit of a sphere amplitude in string theory gives the corresponding tree-level scattering amplitude of gravity~\cite{Scherk:1971xy,Yoneya:1973ca,Scherk:1974ca}. The relationship can also be captured at the non-linear level by considering the worldsheet sigma model on an arbitrary curved background, composed of a metric $g$, $B$-field, and dilaton $\Phi$.  Maintaining worldsheet conformal invariance requires the vanishing of the worldsheet $\beta$-functionals, which imply the target space fields obey certain equations of motion that at low energies are the Einstein equation together with equations of motion for $B$ and $\Phi$~\cite{AlvarezGaume:1980dk,AlvarezGaume:1981hn,Braaten:1985is,Callan:1985ia}. 

The two ways of obtaining target space field equations are of course different aspects of the same thing. Perturbatively, vertex operators in the worldsheet CFT are infinitesimal deformations of the worldsheet action, and correspond to infinitesimal fluctuations of the background geometry (at least for massless states). In order for a vertex operator to be admissible, the fluctuation it describes must obey the target space field equations, linearized around the background. The linearized field equations arise from the requirement that the vertex operators have the correct anomalous conformal weight, reflecting the fact that the non-linear field equations are the condition for vanishing worldsheet Weyl anomaly.

In either approach, for a generic target space it is prohibitively difficult to write down the exact string equations of motion. Rather, one typically works perturbatively in the string length $\sqrt{\alpha^\prime}$, which governs a derivative expansion in the target space geometry, or equivalently a loop expansion parameter in the worldsheet non-linear sigma model. Higher curvature corrections were first seen from the point of view of the $\alpha^\prime$ expansion of amplitudes in~\cite{Gross:1986iv}, and emerge from the four-loop $\beta$-function of the superstring~\cite{Grisaru:1986px,Grisaru:1986kw}. This infinite series of higher-order corrections play an important role in guaranteeing the excellent high energy behaviour of strings.

\medskip

Recently, a new first-order worldsheet theory has been proposed~\cite{Mason:2013sva,Berkovits:2013xba,Adamo:2013tsa} whose spectrum consists only of the states of (type II) supergravity. This theory is chiral, and may be interpreted either as the infinite tension limit of type II strings\footnote{This interpretation is, at present, only heuristic: while the bosonic portion of this theory can be obtained via a chiral infinite tension limit of the Polyakov action~\cite{Mason:2013sva}, it is not known how the fermionic worldsheet fields can be found from an infinite tension limit of type II string theory. In type II string theory, the two sets of worldsheet fermions have opposite chirality, whereas in this model they are of the \emph{same} chirality.}, or as a complexification of worldline supersymmetric quantum mechanics. There are no massive modes in the theory and correspondingly, at genus zero, $n$-point correlation functions of vertex operators compute $n$-point tree level amplitudes in supergravity \emph{exactly}; there are no $\alpha^\prime$ corrections.

A salient feature of the model of~\cite{Mason:2013sva} is that the supergravity amplitudes appear directly in the remarkable representation discovered by Cachazo, He and Yuan~\cite{Cachazo:2013hca,Cachazo:2013iea}. In particular, the worldsheet theory provides a natural explanation of \emph{why} such amplitudes are supported on the solution set of the `scattering equations'. These equations were known to be closely associated with twistor strings~\cite{Witten:2004cp}, and indeed the chiral worldsheet theory of~\cite{Mason:2013sva,Berkovits:2013xba} is closely related to the twistor string constructions of~\cite{Berkovits:2004jj,Skinner:2013xp,Geyer:2014fka}. Strikingly, they also govern string scattering in the \emph{high} energy, fixed angle regime~\cite{Gross:1987ar}.

The theory in~\cite{Mason:2013sva,Berkovits:2013xba,Adamo:2013tsa} describes maps into flat space-time and computes amplitudes perturbatively around flat space. It is natural to ask if there is a formulation describing maps into curved space-time. Since the theory produces pure supergravity amplitudes when linearized around flat space, the supergravity field equations --- with no $\alpha^\prime$ corrections ---  should be the \emph{exact} conditions for quantum consistency of such a model.

\medskip

This paper provides such a description. We begin in section~\ref{sec:review} by briefly reviewing the worldsheet theory of~\cite{Mason:2013sva}, pointing out its key features. In section~\ref{sec:curved-cl} we present, at the classical level, a generalization of this model describing maps into a curved target space. The key is to generalize the worldsheet current algebra that in the flat space model was responsible for localization on the scattering equations. The appropriate generalization is closely related to the Hamiltonian framework of worldline supersymmetry in supersymmetric quantum mechanics~(\textit{c.f.}, \cite{Blau:1992pm,Frohlich:1993es}). These currents are gauged and, as in flat space, at genus zero it is possible to choose a gauge in which the gauge fields vanish so that the currents disappear from the action. The remaining action is free, opening the possibility of making exact statements about its quantum behaviour. In fact, the action we find is a type of supersymmetric curved $\beta\gamma$-system. The quantum properties of curved $\beta\gamma$-systems have been extensively investigated~\cite{Malikov:1998dw,Malikov:1999,Gorbounov:2001,Nekrasov:2005wg,Witten:2005px,Baulieu:2001fi,BenZvi:2008wv,Frenkel:2006fy,Frenkel:2008vz,Ekstrand:2009zd},  and are rather subtle. In section~\ref{sec:curved-q} we examine the behaviour of the currents under diffeomorphisms of both the target space and worldsheet. We learn that the classical curved space currents of section~\ref{sec:curved-cl} acquire quantum corrections. Finally, in section~\ref{sec:anomaly} we show that the algebra generated by the quantum-corrected currents is anomaly free if and only if the target space satisfies the nonlinear supergravity equations of motion, with no higher curvature corrections.

\section{The flat space model}
\label{sec:review}

We begin by briefly reviewing the model of~\cite{Mason:2013sva} that describes gravity perturbatively around flat space. In conformal gauge the worldsheet action is given by
\be
	S = \frac{1}{2\pi}\int_\Sigma P_\mu\delbar X^\mu +\bpsi_{\mu}\dbar\psi^{\mu} + \bar\chi\, \psi^\mu P_\mu + \chi\, \eta^{\mu\nu}\bpsi_\mu P_\nu
+ \frac{e}{2}\, \eta^{\mu\nu}P_\mu P_\nu \,,
\label{action}
\ee
where $P_\mu\in\Omega^{0}(\Sigma,K_\Sigma)$ is a (1,0)-form on $\Sigma$, while $\psi^{\mu}$ is a complex fermion\footnote{In~\cite{Mason:2013sva} $\psi^\mu$ was written in terms of two Majorana fermions $\psi^\mu_{1,2}$ as $\psi_1^{\mu}=\frac{1}{2}(\psi^{\mu}+\eta^{\mu\nu}\bpsi_{\nu})$ and $\psi_2^\mu = \frac{1}{2\im}(\psi^{\mu}-\eta^{\mu\nu}\bpsi_{\nu})$. We have combined $\psi_{1,2}$ into $\psi$ for later convenience.} taking values in $\Pi\Omega^0(\Sigma,K_\Sigma^{1/2})$. The field $e\in\Omega^{0,1}(\Sigma,T_\Sigma)$ behaves like a Beltrami differential and acts as a Lagrange multiplier imposing the constraint $\cH^0:= \eta^{\mu\nu}P_\mu P_\nu=0$. Likewise, the fermionic fields $\bar\chi,\,\chi\in\Omega^{0,1}(\Sigma,T_\Sigma^{1/2})$ enforce the vanishing of $\G^0:=\psi^\mu P_\mu$ and $\bG^0:=\eta^{\mu\nu}\bpsi_\mu P_\nu$. Thus the action is just a chiral generalization of the worldline action for a massless particle with spin.  The terms involving only bosonic fields can also be obtained from a first-order action for standard string theory by taking a chiral $\alpha^\prime\rightarrow 0$ limit~\cite{Mason:2013sva}, so it is tempting to interpret the theory as a chiral, infinite tension limit of the RNS string. See~\cite{Berkovits:2013xba} for the pure spinor version of this model.

The constraint imposed by $e$ is conjugate to the gauge transformation
\be
	\delta X^\mu = \alpha\, \eta^{\mu\nu}P_{\nu}\,, \qquad \delta e= -\delbar \alpha
\label{flatgauge1}
\ee
where $\alpha\in\Omega^0(\Sigma,T_\Sigma)$, while those imposed by $\bar\chi$ and $\chi$ are conjugate to 
\be
	\delta X^\mu = \bar\epsilon\, \psi^\mu\,,\qquad \delta \bpsi_\mu =  \bar\epsilon\, P_\mu\,,\qquad \delta\bar\chi = -\delbar\bar\epsilon
\label{flatgauge2}\ee
and
\be
	\delta X^\mu = \epsilon\,\eta^{\mu\nu}\bpsi_\nu\,,\qquad \delta \psi^\mu = \epsilon\,\eta^{\mu\nu} P_\nu\,,\qquad \delta\chi = -\delbar\epsilon\,,
\label{flatgauge3}
\ee
respectively, where the fermionic parameters $\bar\epsilon$ and $\epsilon$ are valued in $\Omega^{0}(\Sigma, T_{\Sigma}^{1/2})$. All other fields remain invariant in each case. The currents generating these transformations obey the OPEs
\be
	\G^0(z)\, \bG^0(w) \sim \frac{\cH^0}{z-w}\,,\qquad \G^0(z)\, \G^0(w) \sim 0\,,\qquad \bG^0(z)\, \bG^0(w) \sim 0\,,
\label{flat-alg}
\ee
which may be viewed as an infinite tension limit of the standard $\cN=(1,0)$ SUSY algebra. Note that the target space metric enters the model only through $\cH^0$ and $\bG^0$. 

We can use the BRST procedure to fix the gauge redundancies~\eqref{flatgauge1}-\eqref{flatgauge3} as usual. In the absence of vertex operators, at genus zero we can fix a gauge in which $e$, $\bar\chi$ and $\chi$ all vanish. In this gauge, the currents disappear from the action, which becomes free
\be\label{fS}
	S= \frac{1}{2\pi}\int_\Sigma P_\mu\delbar X^\mu +\bpsi_{\mu}\dbar\psi^{\mu} + b\delbar c + \tilde b\delbar \tilde c + \beta\delbar\gamma + \bar\beta\delbar\bar\gamma\,,
\ee
where $c$ is the usual ghost for holomorphic diffeomorphisms of the worldsheet, while $\tilde c$, $\bar\gamma$ and $\gamma$ are ghosts for the transformations~\eqref{flatgauge1}-\eqref{flatgauge3}, respectively.   Note that all the ghost and matter fields are purely left-moving. The BRST operator is
\be\label{fBRST}
Q  =\oint c\,T^{\mathrm{m}}+:bc\,\partial c:+\bar\gamma\,\G^{0}+\gamma\,\bG^{0}+\frac{\tilde{c}}{2}\,\cH^{0}\,,
\ee
where $T^{\mathrm{m}}$ is the holomorphic stress tensor for all matter and ghost systems other than $b\dbar c$ following from~\eqref{fS}. $Q$ is nilpotent provided the space-time dimension is ten, as in the usual superstring. 

As explained in~\cite{Mason:2013sva}, NS-NS sector\footnote{With respect to $\psi_{1,2}$.} vertex operators in this theory correspond to perturbations of the target space metric, $B$-field and dilaton, while in~\cite{Adamo:2013tsa} the NS-R, R-NS and R-R sectors were shown to provide the gravitini and $p$-form fields that complete the spectrum to linearized Type II (A or B) supergravity in ten dimensions.  An important point is that the linearized field equations on the target space emerge from double contractions between the vertex operators and the currents $\G^0$, $\bG^0$ and $\cH^0$ in the BRST operator, rather than from requiring that they have the correct anomalous conformal weight (as would be the case in usual string theory). Indeed, since the $XX$ OPE is trivial, the plane wave $\e^{\im k\cdot X}$ always has vanishing conformal weight, irrespective of the external momentum $k_\mu$. Closely related to this is the fact that the spectrum contains no massive excitations, which have effectively decoupled in the infinite tension limit. 

\medskip

The main claim to fame of this description of supergravity is that it provides the origin of the striking formul\ae\ of Cachazo, He and Yuan~\cite{Cachazo:2013hca,Cachazo:2013iea} for tree-level scattering amplitudes involving arbitrarily many NS sector states. In particular, localization onto solutions of the scattering equations arises as a consequence of the constraint that $\cH^{0}=P^2(z)=0$ identically over $\Sigma$. In the absence of vertex operators, the $X$ path integral forces $P_\mu$ to be holomorphic over $\Sigma$, and hence both $P_\mu$ and $P^2$ vanish automatically at genus zero. When vertex operators are inserted, $P_{\mu}(z)$ becomes meromorphic with simple poles at the insertion points, with $P^2$  likewise becoming a meromorphic quadratic differential.  The finite ($n-3$)-dimensional system of equations
\be\label{fSeq}
	\sum_{j\neq i}\,\frac{k_i\cdot k_j}{z_i-z_j}=0\,, \qquad i\in\{1,\ldots,n-3\}
\ee
fix the moduli of this pointed curve in terms of the external momenta $\{k_i\}$ and just suffice to ensure that $P^2$ again vanishes everywhere. The worldsheet correlator of the $n$ vertex operators can be simply computed (see~\cite{Mason:2013sva,Adamo:2013tsa} for details) and leads directly to the formul\ae\ of~\cite{Cachazo:2013hca}. 

In~\cite{Adamo:2013tsa} we went on to provide a similar formula for $n$-particle scattering of NS states at genus one. The loop integral arises as the integral over the zero mode of the $P_\mu$ fields at genus one, and is expected to diverge. We gave arguments supporting the interpretation of this expression as the one loop integrand of supergravity. We further provided a generalization of the scattering equations~\eqref{fSeq} valid at arbitrary genus $g$. In summary, replacing the world{\it line} description of supergravity by a chiral world{\it sheet} description allows one to trade the problem of computing amplitudes by summing over all graph topologies for the problem of finding solutions of the scattering equations.


\section{Curved target space: classical aspects}
\label{sec:curved-cl}

We now seek to generalize the flat space model to the case of a curved target space. In this section we confine ourselves to a discussion of the action and currents at the classical level. The quantum mechanical behaviour of this model is non-trivial and will be investigated in the following section.

\medskip

Let $(M_\R,g_\R)$ be a pseudo-Riemannian space-time and $(M,g)$ its complexification with holomorphic metric $g$. That is, $g: {\rm Sym}^2\, T_M\to \C$ where $T_M$ is the holomorphic tangent bundle of $M$. Temporarily ignoring the gauge fields $(\chi,\bar\chi, e)$, the natural generalization of the matter action for the case of a curved $(M,g)$ is
\be\label{C1}
	S_{\rm cl}=\frac{1}{2\pi}\int_{\Sigma}P_\mu\dbar X^\mu +\bpsi_\mu\bar D\psi^\mu\ ,
\ee
where the fermions $\psi$ and $\bar\psi$ are now understood to take values in the pullbacks $X^*T_M$ and $X^*T^*_M$, respectively,  while $\bar D\psi^\mu  = \delbar\psi^\mu + \Gamma^\mu_{\nu\rho}\psi^\nu \delbar X^\rho$ is the $(0,1)$-part of the pullback to $\Sigma$ of the Levi-Civita connection on $M$. Notice that, unlike the standard string, here it is not possible to include a four-fermion interaction in $S_{\rm cl}$, since all the fermions are left-moving.

We may further simplify the action by introducing the field $\Pi$ as
\be\label{Pi}
\Pi_{\mu}:= P_{\mu}+\Gamma^{\lambda}_{\mu\nu}\,\bpsi_{\lambda}\,\psi^{\nu}\ ,
\ee
whereupon the matter portion of the worldsheet action becomes
\be\label{cl-action}
	S_{\rm cl}=\frac{1}{2\pi}\int_{\Sigma}\Pi_{\mu}\dbar X^{\mu}+\bpsi_{\mu}\dbar \psi^{\mu}\,,
\ee
and does not depend on the choice of target metric $g$. The presence of the Levi-Civita connection in the definition of $\Pi$ is reflected by its non-tensorial transformation 
\be
\Pi_\mu \ \mapsto\ \tilde\Pi_\mu = \frac{\del X^\nu}{\del \tilde X^\mu}\,\Pi_\nu + 
\frac{\del^2 X^\lambda}{\del\tilde X^\mu\del\tilde X^\nu}\frac{\del\tilde X^\nu}{\del X^\sigma}\,\bpsi_\lambda \,\psi^\sigma
\ee
under the diffeomorphism $X^\mu  \mapsto \tilde{X}^{\mu}(X)$ of $M$, so that classically~\eqref{cl-action} remains invariant.

\smallskip

The target space metric does play a role in the curved space generalization of the currents $\G^0$, $\bG^0$ and $\cH^0$. The action~\eqref{cl-action} is invariant under the supersymmetry transformations
\be\label{curved-trans}
\begin{aligned}
	\delta X^{\mu} & =  -\bar\epsilon\,\psi^{\mu}-\epsilon\,g^{\mu\nu}\bpsi_{\nu}\\
	\delta\psi^\mu &= \epsilon\,g^{\mu\nu}(\Pi_\nu -\Gamma^\kappa_{\nu\lambda}\bpsi_\kappa\psi^\lambda) 
	+ \epsilon\, g^{\kappa\nu}\Gamma^\mu_{\nu\lambda}\bpsi_\kappa\psi^\lambda\\
	\delta\bpsi_\mu &= \bar\epsilon\,\Pi_\mu -\epsilon\, g^{\kappa\nu}\Gamma^\lambda_{\mu\nu} \bpsi_\kappa\bpsi_\lambda\\
	\delta\Pi_\mu &= {\epsilon}\,g^{\rho\sigma}\Gamma^{\nu}_{\rho\mu}\left(\bpsi_{\sigma}\Pi_{\nu}+\bpsi_{\nu}\Pi_{\sigma}\right)
	-\frac{\epsilon}{2}\bpsi_{\nu}\bpsi_{\rho}\psi^{\sigma} R^{\nu\rho}_{\:\:\:\:\mu\sigma}\,
\end{aligned}
\ee
with parameters $\epsilon,\, \bar\epsilon \in \Pi\Omega^0(\Sigma,T^{1/2}_\Sigma)$, where $R^\mu_{\ \nu\kappa\lambda}$ is the Riemann curvature of the Levi-Civita connection. At the classical level, these transformations are generated by the Noether currents
\be\label{ccurr}
\begin{aligned}
	\G^{\rm cl}&:=\psi^{\mu}\left(\Pi_{\mu}-\Gamma^\kappa_{\mu\lambda}\bpsi_\kappa\psi^\lambda\right) = \psi^\mu \Pi_\mu\\
	 \bG^{\rm cl}&:=g^{\mu\nu}\bpsi_{\nu}\left(\Pi_{\mu}-\Gamma^{\kappa}_{\mu\lambda}\bpsi_{\kappa}\psi^{\lambda}\right)
\end{aligned}
\ee
where the equality in the first line follows by the symmetry of the Levi-Civita connection. The Poisson brackets of these curved space currents obey the same algebra
\be\label{alg-cl}
\left\{\G^{\rm cl},\bG^{\rm cl}\right\}=\cH^{\rm cl}\ ,\qquad \left\{\G^{\rm cl},\,\G^{\rm cl}\right\}=0\ , \qquad \left\{\bG^{\rm cl},\bG^{\rm cl}\right\}=0
\ee
as in flat space, where now 
\be\label{Hcl}
	\cH^{\rm cl} :=  g^{\mu\nu}(\Pi_\mu - \Gamma^\kappa_{\mu\lambda} \bpsi_\kappa\psi^\lambda)(\Pi_\nu - \Gamma^\rho_{\nu\sigma}\bpsi_\rho \psi^\sigma) 
	- \frac{1}{2} R^{\kappa\lambda}_{\ \ \ \mu\nu} \bpsi_\kappa\bpsi_\lambda \psi^\mu\psi^\nu\ .
\ee
The currents~\eqref{ccurr} \&~\eqref{Hcl} thus generalize the flat space currents $\G^0$, $\bG^0$ and $\cH^0$. They take a similar form to the worldline supersymmetry currents and Hamiltonian in supersymmetric quantum mechanics. In particular, since $\Pi_\mu$ is canonically conjugate to $X^\mu$ while $J_\mu^{\ \nu} = \bpsi_\mu \psi^\nu$ generates target space Lorentz transformations, after quantization $\cH^{\rm cl}$ is a Lichnerowicz Laplacian acting on forms on the infinite dimensional space of maps from $\Sigma$ to $M$.

\smallskip

If there is a $B$-field on $M$, with 3-form field strength $H=\rd B$, then the currents are further modified to 
\be
\begin{aligned}\label{ccon}
\G^{\rm cl} &= \psi^\mu\Pi_\mu + \frac{1}{3!} \psi^\mu\psi^\nu\psi^\kappa\,H_{\mu\nu\kappa}\\
\bG^{\rm cl} &= g^{\mu\nu}\bpsi_{\nu}\left(\Pi_{\mu}-\Gamma^{\kappa}_{\mu\lambda}\bpsi_{\kappa}\psi^{\lambda}\right) 
+ \frac{1}{3!} \bpsi_\mu\bpsi_\nu\bpsi_\kappa H^{\mu\nu\kappa}\\
\cH^{\rm cl}& =g^{\mu\nu}\left(\Pi_{\mu}-\Gamma^{\kappa}_{\mu\lambda}\bpsi_{\kappa}\psi^{\lambda} + \frac{1}{2}H_{\mu\kappa\lambda}\psi^\kappa\psi^\lambda\right)
\left(\Pi_{\nu}-\Gamma^{\rho}_{\nu\sigma}\bpsi_{\rho}\psi^{\sigma}+\frac{1}{2}H_{\nu\rho\sigma}\bpsi^\rho\bpsi^\sigma\right)\\
&\hspace{1cm} - \frac{1}{2} R^{\kappa\lambda}_{\ \ \ \mu\nu} \bpsi_\kappa\bpsi_\lambda \psi^\mu\psi^\nu
-\frac{1}{3!}\psi^{\mu}\bpsi_{\nu}\bpsi_{\kappa}\bpsi_{\lambda}\,\nabla_{\mu}H^{\nu\kappa\lambda}
-\frac{1}{3!}\bpsi_{\mu}\psi^{\nu}\psi^{\kappa}\psi^{\lambda}\,\nabla^{\mu}H_{\nu\kappa\lambda}\ ,
\end{aligned}
\ee
without changing the action. The Poisson brackets of these currents still obey~\eqref{alg-cl}. Note that the $B$-field here does {\it not} appear simply as the torsion of the connection, but rather breaks the $\C^*$-symmetry of the fermion system to $\Z_2$. As in the classical string, including a target space dilaton is best done in the context of the quantum theory.

At the classical level, the transformations generated by these currents, with local parameters $\{\bar\epsilon,\epsilon,\alpha\}$ respectively, are gauge symmetries of the action
\be
	S = \frac{1}{2\pi}\int_\Sigma \Pi_\mu \delbar X^\mu + \bpsi_\mu\delbar\psi^\mu + \bar\chi\G^{\rm cl} + \chi\bG^{\rm cl} + \frac{e}{2}\cH^{\rm cl}
\ee
provided the gauge fields transform as $\delta \bar\chi = -\delbar\bar\epsilon$, $\delta\chi=-\delbar\epsilon$ and $\delta e = -\delbar\alpha$. As in the flat space model, at genus zero, in the absence of vertex operators it is possible to choose the parameters so that the gauge fields vanish and the currents disappear from the action.


\section{Quantum corrections}
\label{sec:curved-q}

In the previous section we showed that the generalization of the flat space model to a curved target involved changing the currents, but not the kinetic terms in the action. In the gauge where $e$, $\chi$ and $\bar\chi$ vanish the worldsheet action is free and the theory knows about the target space fields $(g,B,\Phi)$ only through the BRST operator.  The resulting action is an example of a \emph{curved} $\beta\gamma$-\emph{system}.

We now examine the properties of this theory at the quantum level. The quantum behaviour of curved $\beta\gamma$-systems is known to be subtle~\cite{Malikov:1998dw,Malikov:1999,Gorbounov:2001,Nekrasov:2005wg,Witten:2005px}, though the supersymmetric case is much more straightforward than the purely bosonic one~\cite{BenZvi:2008wv,Frenkel:2006fy,Frenkel:2008vz,Ekstrand:2009zd}. The first piece of good news is that since the action is free, correlation functions may be computed using the free OPEs
\be\label{fprops}
	X^{\mu}(z)\;\Pi_{\nu}(w)\sim \frac{\delta^{\mu}_{\ \nu}}{z-w}\,,\qquad \psi^{\mu}(z)\;\bpsi_{\nu}(w)\sim\frac{\delta^{\mu}_{\ \nu}}{z-w}\,.
\ee
This is one of the main advantages of curved $\beta\gamma$-systems in general. It also sits harmoniously with the motivation for this paper explained earlier: the curved space version of a worldsheet theory describing pure supergravity --- with no higher curvature corrections --- should be solvable.

In this section we use this OPE to examine the transformation properties of the currents~\eqref{ccon} under diffeomorphisms of both the target and worldsheet. We will see that these currents must receive corrections in order to be covariant at the quantum level.

\subsection{Target space diffeomorphisms}
\label{sec:V}

Infinitesimally, target space diffeomorphisms are generated by the Lie derivative $\cL_V$ along some vector field $V$. To realize this in the quantum theory, we must seek an operator $\cO_V$ that generates this diffeomorphism. In order for $\cO_V$ to represent the diffeomorphism algebra, given two vectors $V$ and $W$ we require $\cO_V$ and $\cO_W$ to have the OPE
\be\label{diffOPE}
 	\cO_V(z)\,\cO_W(w)\sim \frac{\cO_{[V,\,W]}(w)}{z-w}\,,
\ee
where $[V,\,W]$ is the Lie bracket of the two vector fields. One might naively try 
\be
	\cO_V^{\rm naive}(z) := -:\!V^\mu(X)\Pi_\mu\!:\  \equiv\ \lim_{\epsilon\to 0} \left( V^\mu(X(z+\epsilon)) \, \Pi_\mu(z) - \frac{1}{\epsilon} \del_\mu V^\mu(z)\right)\,,
\ee
but this fails for two reasons. Firstly, the OPE $\cO^{\rm naive}_V(z) \,\cO_W^{\rm naive}(w)$ does not agree with~\eqref{diffOPE}  because of double contractions. This is a common feature of curved $\beta\gamma$-systems~\cite{Malikov:1998dw,Malikov:1999,Gorbounov:2001,Nekrasov:2005wg,Witten:2005px} whose resolution usually requires replacing the Lie bracket on $T_M$ by the Courant bracket on $T_M\oplus T^*_M$. In our supersymmetric context a further problem with $\cO_V^{\rm naive}$ is that it does not act on the fermions, whereas these transform non-trivially under Diff$(M)$ as they take values in the pullbacks of the target space tangent and cotangent bundles.

Remarkably, these two problems cure one another. The operator
\be\label{lie}
 \cO_V:=-\left(:\!V^\mu\Pi_\mu\!: + \partial_\nu V^\mu\,:\!\bpsi_\mu\psi^\nu\!:\right)
\ee
both obeys the desired OPE~\eqref{diffOPE} and generates the correct Diff$(M)$ transformations of all fields. That is, we have the OPEs
\be
\begin{aligned}
 &\cO_V(z)\, X^\mu(w)\sim \frac{V^\mu(w)}{z-w}\,, \quad \cO_V(z)\, \psi^\mu(w)\sim\frac{\del_\nu V^\mu\,\psi^\nu(w)}{z-w}\,,
 \quad\cO_V(z)\,\bpsi_\mu(w)\sim \frac{-\del_\mu V^\nu\,\bpsi_\nu(w)}{z-w}\\
&\cO_V(z)\,\Pi_\mu(w) \sim -\frac{1}{z-w}\left(:\!\del_\mu V^\nu\,\Pi_\nu\!: +\del_\mu\del_\nu V^\kappa\,:\!\bpsi_\kappa\psi^\lambda\!:\right)(w)\ ,
\end{aligned}
\ee
where again the second term in the transformation of $\Pi$ is the expected non-tensorial behaviour of the Levi-Civita connection. The fact that supersymmetric curved $\beta\gamma$-systems behave more straightforwardly under Diff$(M)$ than their bosonic counterparts has been noted before, see {\it e.g.}~\cite{BenZvi:2008wv,Frenkel:2006fy,Frenkel:2008vz,Ekstrand:2009zd}.

\smallskip

Although the choice~\eqref{lie} ensures that the fundamental fields $\{X,\Pi,\psi,\bpsi\}$ transform as expected under target space diffeomorphisms, this does not guarantee that the same is true of composite operators because of the potential for double (or higher) contractions between $\cO_V$ and the composite operator. In particular, while at the classical level the currents $\G^{\rm cl}$, $\bG^{\rm cl}$ and $\cH^{\rm cl}$ introduced in~\eqref{ccon} transform geometrically under Diff$(M)$, this is not true in the quantum theory. For example, the OPE of $\G^{\rm cl}$ with $\cO_V$ contains a non-vanishing first-order pole
\be
 \cO_{V}(z)\,\G^{\rm cl}(w)\sim \cdots + \frac{\partial\left(\partial_{\mu}\partial_{\nu}V^{\mu}\,\psi^{\nu}\right)}{z-w}+\cdots\,,
\ee
which does not combine with other terms to form any sort of Lie derivative along $V$. As with all quantum anomalies, the origin of this term is a double contraction between $\G^{\rm cl}$ and $\cO_V$.

To correct this anomalous behaviour, the currents~\eqref{ccon} must be modified in the quantum theory. The required modification is to add new terms that involve (holomorphic) worldsheet derivatives. Such terms generate both new contributions to the higher-order pole terms in the OPE with $\cO_V$, and also modify the coefficients of the simple poles by terms involving worldsheet derivatives. After some experimentation, one finds that the modifications should be
\be
\begin{aligned}
\label{q-curr}
	\G&=\ :\!\G^{\rm cl}\!:+\,\partial\left(\psi^\mu\Gamma^\kappa_{\mu\kappa}\right)\\
	\bG&=\ :\!\bG^{\rm cl}\!:+\,g^{\mu\nu}\partial\left(\bpsi_\kappa\Gamma^\kappa_{\mu\nu}\right)\,.
\end{aligned}
\ee
These quantum currents do indeed behave appropriately under target space diffeomorphisms, having the OPEs

\be
	\cO_V(z)\,\G(w) \sim  \cdots + \frac{\cL_V\G}{z-w}\,,\qquad\qquad \cO_V(z)\,\bG(w)\sim \cdots +\frac{\cL_V\bG}{z-w}\,,
\ee
and so are covariant under target space diffeomorphisms at the quantum level.  We emphasize that while the net transformation of the currents appears to be due to the new, derivative terms, this calculation involves non-trivial cancellations between terms arising from double contractions from both the classical and quantum parts of $\G$ and $\bG$. 

In order to include a dilaton it is convenient to rewrite these currents as
\be
\begin{aligned}
\label{q-curr-dil}
	\G&=\ :\!\G^{\rm cl}\!:+\,\partial\left(\cL_{\psi^{\mu}\partial_{\mu}}\log\Omega\right)\,,\\
	\bG&=\ :\!\bG^{\rm cl}\!:+\,\partial\left(\cL_{g^{\mu\nu}\bpsi_{\mu}\partial_{\nu}}\log\Omega\right)+\bpsi_{\mu}\Gamma^{\mu}_{\nu\rho}\,\partial g^{\nu\rho}\,,
\end{aligned}
\ee
where $\Omega=X^*(\sqrt{g}\,\rd x^1\wedge\cdots\wedge\rd x^d)$ is the pullback to $\Sigma$ of a top holomorphic form on the (complex) target space $M$.\footnote{Thus, for any vector field $V$, $\cL_V\log\Omega = \Omega^{-1} \cL_V\Omega = \nabla_\mu V^\mu$, where $\nabla_\mu$ is the Levi-Civita covariant derivative. The existence of $\Omega$ is not restrictive on an affine complex space, but may be expected to lead to interesting constraints on possible compactifications.}
To incorporate a dilaton field $\Phi$ on $M$, we simply choose $\Omega$ to be the pullback of $\e^{-2\Phi}\sqrt g\,\rd x^1\wedge\cdots\wedge \rd x^d$ instead.


\subsection{Worldsheet diffeomorphisms}
\label{sec:T}

While the quantum corrections ensure the currents~\eqref{q-curr} transform covariantly under Diff$(M$) transformations, they also affect their behaviour under \emph{worldsheet} diffeomorphisms.  This can be seen by considering their OPEs with the worldsheet stress tensor
\be
	T^{\rm cl} := -:\Pi_\mu\del X^\mu\!: - \frac{1}{2}\left(:\!\bpsi_\mu \del\psi^\mu\!:  +:\!\psi^\mu\del\bpsi_\mu\!:\right) 
\label{cT}
\ee
that follows from the free action~\eqref{cl-action}.  For example, there is now a triple pole in the OPE between the stress tensor and $\G$
\begin{equation*}
 T^{\rm cl}(z)\,\G(w)\sim -\frac{1}{2}\frac{\cL_{\psi^\mu\del_\mu} \log\Omega}{(z-w)^3}+\cdots\,,
\end{equation*}
showing that $\G$ is no longer primary. The resolution is to modify the stress tensor by a total derivative term; that is, we choose the stress tensor of the quantum theory to be
\be\label{qT}
	T:= T^{\rm cl} - \frac{1}{2}\del^2\, \log\left(\e^{-2\Phi}\sqrt {g}\right)\ .
\ee
It is straightforward to check that using this stress tensor, the currents $\G$ and $\bG$ of~\eqref{q-curr} are primary operators, transforming as sections of $K_\Sigma^{3/2}$ under worldsheet diffeomorphisms. 

Note that unlike in string theory, this modification does not affect the condition for worldsheet conformal invariance, because here the $X(z)\,X(w)$ OPE is trivial, so there are no new contributions to the fourth order pole in $T(z)\,T(w)$. Thus, despite the presence of a non-trivial metric, $B$-field and dilaton on the target, the only restriction on the model to emerges from the $T(z)\,T(w)$ OPE (including ghosts) is the critical dimension dim$_\C(M)=10$, as in flat space. In particular, unlike in usual string theory~\cite{Callan:1985ia,Banks:1986fu}, the target space field equations do not appear in $T(z)\,T(w)$, and so are not related to worldsheet $\beta$-functions. This is as expected from the flat space theory~\cite{Mason:2013sva,Adamo:2013tsa} reviewed in section~\ref{sec:review}: the requirement that the vertex operators had to obey linearized field equations came not from any anomalous conformal weight, but rather from their potentially anomalous behaviour under transformations generated by the gauged currents.

The choice~\eqref{qT} of stress tensor implies that the worldsheet action should likewise be modified to 
\be
	S\rightarrow S+\frac{1}{8\pi}\int_{\Sigma} R_{\Sigma}\,\log\left(\e^{-2\Phi}\sqrt{g}\right)\,,
\label{dilact}
\ee
where $R_\Sigma$ is the worldsheet curvature. We can always choose $R_\Sigma$ to vanish locally in two dimensions, so the addition of this term does not affect the short distance OPE, and our calculations are self-consistent. Actually, the dilaton coupling~\eqref{dilact} is well-known in first-order formulations of string theory~\cite{Schwarz:1992te,Nekrasov:2005wg}, in particular the fact that the dilaton is effectively shifted $\Phi\to \Phi -\frac{1}{2}\log \sqrt g$ compared to the usual dilaton coupling in string theory. This shifted coupling also plays an important role in T-duality, see {\it e.g.}~\cite{Buscher:1987qj}, and analogous shifts also appear when studying $\alpha^\prime$-corrections to string theory using doubled geometry ({\it c.f.},~\cite{Hohm:2013jaa}).


\section{Supergravity equations of motion as an anomaly}
\label{sec:anomaly}

In the previous section we constructed currents~\eqref{q-curr} that behave correctly under both target space and worldsheet diffeomorphisms at the quantum level. Contrary to usual string theory, the requirement of quantum worldsheet conformal invariance places no restrictions on the target space fields. 

Instead, the target space field equations arise from quantum consistency of the current algebra. At the quantum level, the Poisson bracket relations
\be\label{alg-cl2}
	\left\{\G^{\rm cl},\bG^{\rm cl}\right\}=\cH^{\rm cl}\ ,\qquad \left\{\G^{\rm cl},\,\G^{\rm cl}\right\}=0\ , \qquad \left\{\bG^{\rm cl},\bG^{\rm cl}\right\}=0
\ee
between the classical currents should be replaced by OPEs of the quantum currents~\eqref{q-curr}, so that the $\G(z)\,\G(w)$ and $\bG(z)\,\bG(w)$ OPEs are non-singular, while the $\G(z)\,\bG(w)$ OPE has only a simple pole. Only if this is true, so that the algebra of currents is non-anomalous, will the BRST operator~\eqref{fBRST} obey $Q^2=0$. It is a remarkable fact that because the worldsheet action is (locally) free, we can compute these current OPEs \emph{exactly}, and so obtain the exact quantum consistency conditions. This is quite distinct from the usual case in string theory, where for generic backgrounds, one is faced with an intractable, interacting worldsheet CFT and so must work perturbatively around some fixed background, treating $\alpha^\prime$ as a loop expansion on the worldsheet, or derivative expansion in the target. We have no $\alpha^\prime$ parameter.

\medskip

We begin with the $\G(z)\,\G(w)$ OPE. Performing all possible contractions and expanding the coefficients of higher order poles around the mid-point, we find
\be\label{GG1}
	\G(z)\,\G(w)\sim -\frac{1}{3}\frac{\psi^{\kappa}\psi^{\lambda}\psi^{\mu}\psi^{\nu}}{z-w}\,\del_{\kappa}H_{\lambda\mu\nu}-\frac{\partial\left(\psi^{\mu}\psi^{\nu}\,\partial_{\mu}\Gamma^{\kappa}_{\nu\kappa}\right)}{z-w}+2\,\frac{\partial\left(\psi^{\mu}\psi^{\nu}\del_{\mu}\del_{\nu}\Phi\right)}{z-w}\, .
\ee
The second and third terms in this expression vanish by the antisymmetry of fermions contracted into partial derivatives. (Recall that $\partial_{\mu}\Gamma^{\kappa}_{\nu\kappa}=\partial_{\mu}\partial_{\nu}\log\sqrt{g}$.)  Hence the only non-trivial anomaly cancellation condition in \eqref{GG1} is given by the first term. This is simply the requirement that the 3-form $H$ is closed so that $H=\rd B$ at least locally on $M$. Thus $H$ is indeed the field strength of a $B$-field. 

We now turn to the $\bG(z)\,\bG(w)$ OPE. Again performing all possible contractions and expanding around the mid-point we find
\be\label{bGbG1}
\begin{aligned}
 \bG(z)\;\bG(w)\ \sim\ &\frac{1}{2}\frac{:\!\bpsi_{\kappa}\bpsi_{\lambda}\bpsi_{\mu}\psi^{\nu}\!:}{z-w}\,R_{\nu}^{\ \, \kappa\lambda\mu}
 +\frac{\del\left(\bpsi_{\mu}\bpsi_{\nu}\,R^{\mu\nu}\right)}{z-w} -\frac{1}{3}\frac{\bpsi_{\kappa}\bpsi_{\lambda}\bpsi_{\mu}\bpsi_{\nu}}{z-w}\,\partial^{\kappa}H^{\lambda\mu\nu} \\
&\ \ +2\frac{\bpsi_{\mu}\bpsi_{\nu}\,\partial X^{\kappa}}{z-w}\left[\Gamma^{\nu}_{\rho\sigma} \,R^{\sigma\rho\mu}_{\ \ \ \ \kappa}+\Gamma^{\rho}_{\kappa\sigma}(R^{\mu\sigma\nu}_{\ \ \ \ \rho}+R^{\nu\sigma\mu}_{\ \ \ \ \rho})\right]\ .
\end{aligned}
\ee
These anomalies vanish provided again $\rd H=0$ and the Riemann and Ricci tensors obey the identities
\be\label{A2}
R_{\nu}^{\:\:[\kappa\lambda\mu]}=0\,, \quad R^{[\mu\nu]}=0\quad\hbox{and}\quad R^{(\mu\nu)}_{\:\:\:\:\:\:\rho\sigma}=0\, .
\ee
These are of course the first Bianchi identity and basic symmetries of the Riemann and Ricci tensors that hold provided the connection $\Gamma$ is indeed Levi-Civita. So neither of these two OPEs impose any dynamical restrictions on the target space fields.

The only remaining OPE to be checked is that of $\G(z)$ and $\bG(w)$.  This OPE has first, second and third order poles. The coefficient of the first order pole defines the quantum corrected current $\cH$, but the coefficients of the higher order poles must be made to vanish. Proceeding as above, a straightforward, if somewhat lengthy, calculation yields
\be\label{GbG1}
\begin{aligned}
 \G(z)\,\bG(w)\ \sim \ \ &\frac{2}{(z-w)^3}\left(R+4\nabla_{\mu}\nabla^{\mu}\Phi-4\nabla_{\mu}\Phi\,\nabla^{\mu}\Phi-\frac{1}{12}H^{2}\right) \\
& \ \ \ +2\,\frac{(\Gamma^{\mu}_{\kappa\nu}\partial X^{\kappa}+\psi^{\mu}\bpsi_{\nu})}{(z-w)^2}\,g^{\nu\lambda}\left(R_{\mu\lambda}+2\nabla_{\mu}\nabla_{\lambda}\Phi-\frac{1}{4}H_{\mu\rho\sigma}H_{\lambda}^{\ \rho\sigma}\right) \\
&\ \ \  +\frac{(\psi^{\mu}\psi^{\nu}-\bpsi^{\mu}\bpsi^{\nu})}{(z-w)^2}\left(\nabla_{\kappa}H^{\kappa}_{\ \mu\nu}-2H^{\kappa}_{\  \mu\nu}\nabla_{\kappa}\Phi\right) +\frac{\cH}{z-w}\,.
\end{aligned}
\ee
The quantum corrected current $\cH$ takes the somewhat unenlightening form
\be
\begin{aligned}
	\cH=\cH^{\rm cl}  +\partial\left(\cL_{g^{\mu\nu}\Pi_\mu\del_\nu}\log\Omega\right) -\frac{1}{2}\partial^{2}(g^{\mu\nu})\,\partial_{\mu}\partial_{\nu}\log\left(\sqrt{g}\e^{-2\Phi}\right)-\bpsi_{\kappa}\partial\psi^{\lambda}\,g^{\mu\nu}\partial_{\lambda}\Gamma^{\kappa}_{\mu\nu} \\
-\frac{1}{4}\partial(g^{\mu\nu})\,\partial\left[\partial_{\mu}\partial_{\nu}\log\left(\sqrt{g}\e^{-2\Phi}\right)\right]+\frac{1}{2}H^{\mu\nu\kappa}\bpsi_{\kappa}\,\partial\left(H_{\mu\nu\lambda}\psi^{\lambda}\right) +\partial\left(H_{\kappa\lambda\nu}\psi^{\nu}\right)\,g^{\kappa\sigma}\Gamma^{\lambda}_{\sigma\rho}\psi^{\rho}\\
-\frac{1}{2}\partial_{\sigma}H_{\mu\nu\rho}\psi^{\nu}\psi^{\rho}\,\partial(g^{\sigma\mu})
-\frac{1}{12}H^{\mu\nu\rho}\partial^{2}H_{\mu\nu\rho}+\frac{1}{2}\partial(g^{\mu\nu})\Gamma^{\rho}_{\mu\nu}\,\left(2\Pi_{\rho}+H_{\sigma\lambda\rho}\psi^{\sigma}\psi^{\lambda}\right) \\
-\partial\left[\partial(g^{\mu\nu})\left(\partial_{\sigma}\Phi\Gamma^{\sigma}_{\mu\nu}+\frac{1}{2}\Gamma^{\sigma}_{\mu\nu}\Gamma^{\rho}_{\sigma\rho}-\frac{1}{2}\partial_{\sigma}\Gamma^{\sigma}_{\mu\nu}\right)+g^{\mu\nu}\Gamma^{\rho}_{\mu\sigma}\partial(\Gamma^{\sigma}_{\nu\rho})\right] \\
-\partial\left[\bpsi_{\kappa}\psi^{\lambda}\left(\nabla^{\kappa}\nabla_{\lambda}\Phi-2g^{\mu\nu}\Gamma^{\kappa}_{\mu\lambda}\partial_{\nu}\Phi\right)\right]\,. 
\end{aligned}
\ee
From~\eqref{GbG1} we see that the algebra of currents is anomaly free if and only if the space-time fields $(g,B,\Phi)$ obey the equations
\be\label{eom}
\begin{aligned}
 R_{\mu\nu}-\frac{1}{4}H_{\mu\kappa\lambda}\,H_{\nu}^{\ \kappa\lambda}+2\nabla_{\mu}\nabla_{\nu}\Phi\ &=&  0\,,\\
 \nabla_{\kappa}H^{\kappa}_{\ \mu\nu}-2H^{\kappa}_{\ \mu\nu}\nabla_{\kappa}\Phi\ &=& 0\,, \\
 R+4\nabla_{\mu}\nabla^{\mu}\Phi-4\nabla_{\mu}\Phi\,\nabla^{\mu}\Phi-\frac{1}{12}H^2\ & =&  0\,.
\end{aligned}
\ee
These are precisely the field equations of general relativity with a $B$-field and dilaton.  Hence, the \emph{exact} condition for the worldsheet theory to be consistent at the quantum level is that the target space $(M;g,B,\Phi)$ obeys the non-linear $d=10$ supergravity field equations, in the Neveu-Schwarz sector.  

The BRST operator constrains physical field configurations to obey $\cH=0$, which in flat space is the condition $\eta^{\mu\nu}\Pi_\mu\Pi_\nu=0$ at every point of the worldsheet. As reviewed in section~\ref{sec:review}, this the content of the scattering equations. The $\G(z)\,\bG(w)$ OPE has $\cH$ as its classical contribution, while the field equations~\eqref{eom} appear as the coefficients of higher poles. In this sense, the Einstein equations emerge as quantum corrections to the curved space generalization of the scattering equations.


\section{Conclusions}
\label{sec:conc}

In this paper we have constructed a worldsheet theory that describes maps to a (complexified) Riemannian manifold $(M,g)$. The worldsheet action is a type of supersymmetric curved $\beta\gamma$-system. Unlike for a purely bosonic curved $\beta\gamma$-system, the path integral has no anomalous behaviour under diffeomorphisms of $M$, and so is free. The theory involves gauging a certain worldsheet superalgebra which generates supersymmetry-like transformations, and is a chiral analogue of the supersymmetry algebra in string theory. The corresponding algebra of currents is anomaly free, and hence the BRST operator is nilpotent, if and only if the target space obeys the $d=10$ supergravity equations of motion, with no higher curvature corrections.

\medskip

We close with a few remarks. Firstly, the curved space worldsheet theory encodes the vertex operators for perturbations of the metric, $B$-field, and dilaton around flat space.  In the non-linear sigma model of string theory, the flat space vertex operators are found by considering linearized perturbations of the action; here the vertex operators arise by perturbing the \emph{currents}.  For example, expanding the metric in $\cH$ to linear order around the Minkowski metric one finds
\be	
	\cH -\cH^0 =  \delta g^{\mu\nu}\,\Pi_\mu\Pi_\nu - 2 \eta^{\mu\nu}\,\Pi_\mu \delta\Gamma^\kappa_{\nu\lambda}\,\bar\psi_\kappa\psi^\lambda 
	 -\del_\mu(\delta\Gamma^\kappa_{\nu\lambda})\,\bpsi_\kappa\bpsi^\lambda\psi^\mu\psi^\nu, 
\ee
up to terms which vanish on the support of the flat space scattering equations $\cH^0=\eta^{\mu\nu}\Pi_{\mu}\Pi_{\nu}=0$.  This quadratic differential is essentially the vertex operator describing fluctuations $\delta g$ around flat space. When the fluctuations are plane waves with target space momentum $k_\mu$, the remaining factor of  the integrated vertex operator is $\bar\delta(k\cdot\Pi)\in H^{0,1}(\Sigma,T_\Sigma)$, which is best interpreted as a modulus of the gauge field $e$ on the marked worldsheet.  The integrated vertex operators describing fluctuations  $\delta B$ or $\delta\Phi$ around flat space are obtained similarly. Expanding the currents $\G$ and $\bG$ around flat space (and re-expressing them in terms of real fermions) likewise gives the vertex operators in different `pictures'. See~\cite{Mason:2013sva} for details.

Secondly, note that the dilaton equation of motion enters in the $\G(z)\,\bG(w)$ OPE~\eqref{GbG1} at order $(z-w)^{-3}$, whereas  the Einstein and $B$-field equations enter at order $(z-w)^{-2}$. This is analogous to the way the dilaton equation of motion appears at higher loop order in the worldsheet $\beta$-functionals in usual string theory. Of course, the dilaton equation of motion is implied by the Einstein and $B$-field equations, so that the triple pole in $\G(z)\,\bG(w)$ is guaranteed to vanish if the double poles do. In this sense, the exact target space field equations indeed arise from a 1-loop anomaly of the currents.

\smallskip

We briefly consider dimensional reduction of the theory presented here. Perturbatively, we expect that this should correspond to Kaluza-Klein reductions of $d=10$ supergravity. Amplitudes involving scattering of Kaluza-Klein excitations provide a natural example of the \emph{massive} scattering equations presented in~\cite{Naculich:2014naa}. Non-perturbatively, we must remember that the theory here lives on a Riemann surface, and there can be worldsheet instantons wrapping holomorphic curves in the target. These effects go beyond what can be seen in a purely worldline description of supergravity. More generally, it would be fascinating to understand whether D-branes can survive in this infinite tension limit, despite the worldsheet theory being chiral.

\smallskip

In this paper we have concentrated on the geometry of $M$, viewing the theory as a complexification of worldline supersymmetric quantum mechanics, or as a chiral infinite tension limit of the superstring. Another perspective is also useful. The fields $(\Pi,X)$ together describe a map from the worldsheet to the (holomorphic) cotangent bundle of $M$ (with $\Pi$ twisted by the worldsheet canonical bundle). Imposing the constraint $\cH=0$ and quotienting by the gauge transformations generated by $\cH$ amounts to taking the symplectic quotient of $T^*_M$ by $\cH$. The resulting space is the space of null rays in $M$, often known as the ambitwistor space of $M$~\cite{LeBrun:1983}. Adding the worldsheet fermions provides a supersymmetric version of this space. Thus the model can also be said to describe an ambitwistor string theory, which was the point of view adopted in~\cite{Mason:2013sva}. 

There are two main advantages to this perspective. Firstly, the vertex operators which here are treated as deformations of the currents $\G$, $\bG$ and $\cH$ are naturally interpreted in terms of cohomology classes on ambitwistor space, and in fact give a simple example of the Penrose transform (see {\it e.g.}~\cite{WardWells}). Secondly, and perhaps more importantly, the ambitwistor space has several other representations. For example, in a globally hyperbolic space-time the space of null rays may be identified with 
the bundle of null directions over any Cauchy surface. An important special case is in asymptotically flat space-times where the Cauchy surface is chosen to limit onto (past or future) null infinity. The relation of the flat space model to descriptions of gravity living at null infinity has been explored in~\cite{Adamo:2014yya,Geyer:2014lca}. It would be interesting to revisit these from the present, curved perspective.

For some space-times it may even be possible to \emph{solve} the constraint $\cH=0$ exactly. For example, in four dimensional flat space-time the constraint $P^2=0$ is solved by writing the momentum as a simple bispinor $P= \lambda\,\tilde\lambda$ and (once compactified and projectivized) the ambitwistor space can be viewed as a quadric hypersurface in $\CP^3\times\CP^3$. It would be fascinating if the curved model here could lead to a deeper understanding of the twistor and ambitwistor models of~\cite{Skinner:2013xp,Geyer:2014fka}.

\smallskip

Throughout this paper we have used the `RNS' formulation of the worldsheet action. This has the advantage that the worldsheet fermionic spinors provide a natural origin of the Pfaffians in the $n$-point CHY amplitude formula~\cite{Cachazo:2013hca}. However, it has the usual disadvantage of the RNS formulation that while vertex operators corresponding to target space gravitinos and form fields can be constructed, it is difficult to understand how to turn on nonlinear background fields beyond the NS sector. In~\cite{Berkovits:2013xba}, Berkovits presented a pure spinor version of the flat space model of section~\ref{sec:review}. It is important to understand how to generalize the pure spinor model to curved backgrounds. This seems particularly important, both in the light of the close relation~\cite{Berkovits:2014aia} between twistors and the pure spinor string at finite $\alpha^\prime$, and in terms of the interpretation of the supergravity field equations as integrability of a certain superconnection along super null rays~\cite{Witten:1985nt}. Indeed, this was one of the original motivations for considering ambitwistor spaces.

Finally, it would clearly be very interesting to understand how the considerations of this paper can be adapted to relate the CHY formula~\cite{Cachazo:2013hca} for scattering of gluons to the (super-)Yang-Mills equations. Finding an (ambi-)twistor string for pure SYM is a long standing problem. 


\acknowledgments

We thank Malcolm Perry and David Tong for helpful discussions. The work of TA is supported by a Title A Research Fellowship at St. John's College, Cambridge. The work of EC is supported in part by the Cambridge Commonwealth, European and International Trust. DS is supported in part by a Marie Curie Career Integration Grant (FP/2007-2013/631289). Research leading to these results has received funding from the European Research Council under the European Community's Seventh Framework Programme (FP7/2007-2013) / ERC grant agreement no. [247252].

\bibliography{Einst}
\bibliographystyle{JHEP}

\end{document}